\documentstyle[aps,floats,epsf,twocolumn]{revtex}
\begin{document} \draft

\title{Quasiparticles in the vortex lattice of unconventional 
superconductors:\\ Bloch waves or Landau levels? }

\author{M. Franz and Z. Te\v{s}anovi\'c}
\address{Department of Physics and Astronomy, Johns Hopkins University,
Baltimore, MD 21218
\\ {\rm(\today)}
}
%
\address{~
\parbox{14cm}{\rm
\medskip
A novel singular gauge transformation is developed for quasiparticles
in the mixed state of a strongly type-II superconductor which permits a full
solution of the problem at low and intermediate fields, $H_{c1}< B\ll H_{c2}$.
For a periodic vortex lattice the natural low-energy quasiparticle states 
are Bloch waves rather than
Landau levels discussed in the recent literature. The new representation
elucidates the physics considerably and 
provides fresh insights into the spectral properties 
of such systems. 
}}
\maketitle


%
\narrowtext

In conventional ($s$-wave) superconductors the single particle excitation 
spectrum is gapped and, consequently,
no quasiparticle states are populated at low 
temperatures. The situation is dramatically 
different in unconventional superconductors which exhibit nodes in the gap. 
These lead to finite
density of fermionic excitations at low energies which then
dominate the low-temperature physics.
Among the known (or suspected) unconventional superconductors are high-$T_c$
copper oxides, organic and heavy fermion superconductors, and the 
recently discovered
Sr$_2$RuO$_4$. Understanding the physics of the low-energy quasiparticles 
in the mixed state of these unconventional superconductors is an unsolved
problem of considerable complexity.
This complexity stems from the fact that (i) in the mixed state
superconductivity coexists with the magnetic field ${\bf B}$ and the 
quasiparticles feel the {\em combined} effects of ${\bf B}$ and the 
spatially varying field of chiral supercurrents, and (ii)  
being composite objects, part electron and part hole, 
quasiparticles
 do not carry a definite charge. The corresponding low-energy theory
therefore poses an entirely new intellectual challenge\cite{lee1}, which is 
simultaneously of considerable practical interest. 

The initial theoretical investigations were based 
on numerical computations\cite{wang1}, semiclassical
approximations\cite{volovik1} and general scaling arguments\cite{sl1}. 
More recently Gorkov and Schrieffer\cite{gs1} made a remarkable prediction
that in a $d_{x^2-y^2}$ superconductor 
at intermediate fields $H_{c1}\ll B\ll H_{c2}$ the quasiparticles
will form Landau levels (LL) with a discrete energy spectrum
\begin{equation}
E_n=\pm\hbar\omega_H\sqrt{n}, \ \ n=0,1,\dots,
\label{gs}
\end{equation}
where $\omega_H=2\sqrt{\omega_c\Delta_0/\hbar}$, with $\omega_c=eB/mc$ being
the cyclotron frequency and $\Delta_0$ the maximum superconducting
gap. Based on a somewhat different reasoning
Anderson\cite{pwa1} later arrived at a similar conclusion and argued that 
LL quantization could explain the anomalous magneto-transport in 
cuprates. Jank\'o\cite{janko1} proposed a direct test of Eq.\ (\ref{gs})
using the scanning tunneling spectroscopy. 

The concept of
LL quantization here is quite different from the one in
conventional superconductors, where nodes in the gap arise as a result 
of the {\em center of mass} motion of pairs in {\em strong}
magnetic fields near $H_{c2}$\cite{dukan1}. 
The physics of Eq.\ (\ref{gs}) is based on the picture of a low-energy
quasiparticle Larmor-precessing in {\em weak} external magnetic field
along an elliptic orbit of constant energy centered at the Dirac 
gap node  in the $k$-space\cite{sl1,pwa1}. This motion corresponds to a 
closed elliptic orbit in the real space and the 
quantization condition (\ref{gs}) follows from demanding that such an 
orbit contains $n$ quanta of magnetic flux. This argumentation neglects the
effect of spatially varying supercurrents in the vortex array, which were, 
however, recently shown by Melnikov\cite{melnikov1} to strongly mix
the individual Landau levels. 

In this paper we formulate a new approach to the problem which treats 
the effects of the magnetic field and supercurrents on equal footing. 
As a result the physics becomes transparent and  
for periodic vortex arrays in the intermediate field 
regime the low-energy theory can be solved in its entirety. 
Our principal result is that the natural low-energy quasiparticle states are 
Bloch waves of massless Dirac fermions
and not the Landau levels discussed above. 

At the heart of our
approach is the observation that the collective response of the 
condensate to the external magnetic field on average {\em exactly compensates} 
its effect on the normal quasiparticles. 
More formally, the phase of the superconducting
order parameter, $\Delta({\bf r})=\Delta_0e^{i\phi({\bf r})}$, acts as an 
additional
``gauge field'' coupled to the quasiparticles. In the vortex state 
$\phi({\bf r})$
is {\em not} a pure gauge: $\nabla\times\nabla\phi({\bf r})=
2\pi\hat z\sum_i\delta({\bf r}-{\bf R}_i)$ where $\{{\bf R}_i\}$ denotes 
vortex positions. From the vantage point of a 
quasiparticle the singularities in $\nabla\times\nabla\phi$ 
act as magnetic {\em half-fluxes} 
concentrated in the vortex cores with polarity opposing the external field.  
Flux quantization ensures that {\em on average} this ``spiked'' field
exactly cancels out the external applied field ${\bf B}$. In the mixed state,
the quasiparticle therefore can be thought of as 
moving in an {\em effective} field ${\bf B}_{\rm eff}$ which is zero 
on average 
but derives from a vector potential that is highly nontrivial. The nature of
the phenomenon is  purely quantum-mechanical, and is closely related to the 
Aharonov-Bohm effect: 
classical charged particle would be completely unaffected by the spiked 
field because the singularities occupy a set of measure zero in the 
real space.

Our solution consists in finding a gauge in which the Hamiltonian 
manifestly displays the physical property described above. In such a gauge the 
fermionic excitation spectrum can be found by band structure 
techniques suitably adjusted to the ``off-diagonal''
structure of the theory.  
Besides revealing the nature of the low-energy quasiparticles, this
representation leads to new insights into the physics of the mixed
state. One surprising finding is that in a perfectly periodic vortex lattice
the original Dirac nodes survive the perturbing effect of a weak magnetic 
field.

We now supply the details. Quasiparticle wavefunction
$\Psi^T({\bf r})=[u({\bf r}),v({\bf r})]$ is subject to the 
Bogoliubov-de Gennes equation ${\cal H}\Psi=E\Psi$, where
\begin{equation}
{\cal H} =\left( \begin{array}{cc}
\hat{\cal H}_e      & \hat{\Delta} \\
\hat{\Delta}^*  & -\hat{\cal H}_e^*
\end{array} \right)
\label{h1}
\end{equation}
with $\hat{\cal H}_e={1\over 2m}({\bf p}-{e\over c}{\bf A})^2-\epsilon_F$ 
and $\hat\Delta$
the $d$-wave pairing operator. Following Simon and Lee\cite{sl1} we choose
the coordinate axes in the direction of gap nodes, in which case 
$\hat\Delta=p_F^{-2}\{\hat p_x,\{\hat p_y,\Delta({\bf r})\}\}$, where $p_F$ is
the Fermi momentum, $\hat{\bf p}=-i\hbar\nabla$, and curly brackets represent
symmetrization, $\{a,b\}={1\over 2}(ab+ba)$. In high-$T_c$ cuprates it is
natural to concentrate on the low to 
intermediate field regime  $H_{c1} < B\ll H_{c2}$, where the vortex
spacing is large and we may assume the gap amplitude to be constant 
everywhere in space, 
$\Delta({\bf r})\simeq\Delta_0e^{i\phi({\bf r})}$.
Under such conditions the magnetic field distribution is described by a simple
London model  and superfluid velocity, defined as 
${\bf v}_s({\bf r})={1\over m}({\hbar\over 2}\nabla\phi-{e\over c}{\bf A})$, 
can be
written in terms of  vortex positions $\{{\bf R}_i\}$ as\cite{tinkham}
\begin{equation}
{\bf v}_s({\bf r})={\pi\hbar\over m}\int {d^2k\over(2\pi)^2}
{i{\bf k}\times\hat z\over \lambda^{-2}+k^2}\sum_ie^{i{\bf k}\cdot({\bf r}-
{\bf R}_i)},
\label{vs}
\end{equation}
where $\lambda$ is the London penetration depth. 

In order to diagonalize (\ref{h1}) it is desirable to remove the phase
factors $e^{i\phi({\bf r})}$ from the off-diagonal components of ${\cal H}$. 
This is accomplished by a unitary transformation
\begin{equation}
{\cal H} \to U^{-1}{\cal H}U, \ \ \ 
U=\left( \begin{array}{cc}
e^{i\phi_e({\bf r})}      & 0 \\
0  & e^{-i\phi_h({\bf r})} 
\end{array} \right),
\label{u1}
\end{equation}
where $\phi_e({\bf r})$ and $\phi_h({\bf r})$ are arbitrary functions 
satisfying
\begin{equation}
\phi_e({\bf r})+\phi_h({\bf r})=\phi({\bf r}).
\label{con1}
\end{equation}
Eq.\ (\ref{u1}) can be thought of as a singular gauge transformation since 
it changes the effective magnetic field seen by electrons and holes. We now
discuss three specific choices for the  functions $\phi_e$ and $\phi_h$.

The most natural choice satisfying (\ref{con1}) is the symmetric one, 
namely $\phi_e({\bf r})=\phi_h({\bf r})=\phi({\bf r})/2$, resulting in
\cite{remark4} 
\[
{\cal H}_{S} =\left( \begin{array}{cc}
{1\over 2m}(\hat{\bf p}+m{\bf v}_s)^2-\epsilon_F & {\Delta_0\over p_F^2}
\hat p_x\hat p_y \\
{\Delta_0\over p_F^2}\hat p_x\hat p_y  & 
-{1\over 2m}(\hat{\bf p}-m{\bf v}_s)^2+\epsilon_F
\end{array} \right).
\]
This particular gauge makes the Hamiltonian very simple but unfortunately
is not very useful because, as noted by Anderson\cite{pwa1} and in a  
different context by Balents {\em et al.}\cite{balents1}, the corresponding 
transformation (\ref{u1}) is not single valued. To see this, consider
the situation on encircling the core of a vortex:
$\phi$ winds by $2\pi$ but $\phi_e$ and $\phi_h$
pick only a phase of $\pi$, causing $U$ to have two branches. 
Consequently, one is forced to diagonalize ${\cal H}_S$ under the 
constraint that the {\em original} wavefunctions are single valued.
Clearly, this is a difficult task. Nevertheless, the symmetric gauge 
reveals the physical essence of the problem:
formally, ${\bf v}_s$ enters ${\cal H}_S$ as an {\em effective} vector
potential, which corresponds to an effective magnetic field
${\bf B}_{\rm eff}=-{mc\over e}(\nabla\times{\bf v}_s)\neq {\bf B}$. It is 
easy to 
show from Eq.\ (\ref{vs}) that ${\bf B}_{\rm eff}$ vanishes on average
\cite{remark2}, i.e. that
$\langle\nabla\times{\bf v}_s\rangle=0$, where angular brackets denote the
spatial average.
Aside from the single-valuedness problem, the low-energy 
physics described by ${\cal H}_S$
is that of a quasiparticle in {\em zero average magnetic field}.
The external field is compensated by the array of magnetic half-fluxes
giving rise to a
non-trivial vector potential with the periodicity of the vortex lattice. 

To avoid the problem of multiple valuedness Anderson\cite{pwa1}, suggested 
taking $\phi_e({\bf r})=\phi({\bf r})$ and $\phi_h({\bf r})=0$. This 
leads to a Hamiltonian of the form\cite{remark1}
\[
{\cal H}_{A} =\left( \begin{array}{cc}
{1\over 2m}(\hat{\bf p}+{e\over c}{\bf A}+2m{\bf v}_s)^2-\epsilon_F & \hat D \\
\hat D  & 
-{1\over 2m}(\hat{\bf p}+{e\over c}{\bf A})^2+\epsilon_F
\end{array} \right),
\]
with $\hat D={\Delta_0\over p_F^2}(\hat p_x+{e\over c}A_x+mv_{sx})
(\hat p_y+{e\over c}A_y+mv_{sy})$. In this representation one could 
consider neglecting in the first approximation the ${\bf v}_s$ terms, on the
grounds that they represent a perturbative correction\cite{pwa1}. 
At low energies, 
expanding all the terms in ${\cal H}_A$ to leading order near the nodes, 
the Hamiltonian becomes that of massless Dirac fermions in a uniform 
magnetic field ${\bf B}=\nabla\times{\bf A}$ with the energy spectrum 
\cite{pwa1}
given by Eq.\ (\ref{gs}). The effects of supercurrent field ${\bf v}_s$ 
can in principle be
treated perturbatively. This is, however, difficult in practice, because of
the massive degeneracy of the Landau levels and also because 
${\bf v}_s({\bf r})$
is not a small perturbation (it diverges as $1/r$ at the vortex cores) and
will lead to strong LL mixing\cite{melnikov1}. 
In the absence of a reliable scheme to incorporate ${\bf v}_s({\bf r})$, the 
physical picture of Larmor precessing quasiparticle that leads to 
Eq.\ (\ref{gs}) appears incomplete. 

We now introduce a new singular gauge transformation that combines the
desirable features of the two transformations discussed above but has 
none of their drawbacks. Consider dividing vortices into two distinct subsets
$A$ and $B$, each containing an equal number of vortices.
Now denote by $\phi_A({\bf r})$ the phase field associated with vortices
in the subset $A$, with the analogous definition of $\phi_B({\bf r})$. The 
choice  $\phi_e({\bf r})=\phi_A({\bf r})$, $\phi_h({\bf r})=\phi_B({\bf r})$
clearly satisfies the condition (\ref{con1}) and the corresponding 
transformation $U$ is single valued. The resulting Hamiltonian is
\[
{\cal H}_{N} =\left( \begin{array}{cc}
{1\over 2m}(\hat{\bf p}+m{\bf v}_s^A)^2-\epsilon_F & \hat D \\
\hat D  & 
-{1\over 2m}(\hat{\bf p}-m{\bf v}_s^B)^2+\epsilon_F
\end{array} \right),
\]
with $\hat D={\Delta_0\over p_F^2}[\hat p_x+{m\over 2}(v_{sx}^A-v_{sx}^B)]
[\hat p_y+{m\over 2}(v_{sy}^A-v_{sy}^B)]$ and 
\begin{equation}
{\bf v}_s^\mu={1\over m}(\hbar\nabla\phi_\mu-{e\over c}{\bf A}), \ \ \mu=A,B.
\label{vsab}
\end{equation}
As long as the physical field ${\bf B}$ remains approximately uniform,
\begin{figure}[t]
\epsfxsize=8.5cm
\epsffile{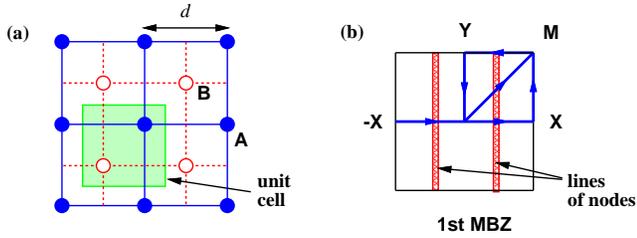}
\caption[]{(a) Two sublattices $A$ and $B$ of the square vortex lattice and 
the unit cell used in the numerical diagonalization of Eq.(\ref{bloch}). (b)
The corresponding magnetic Brillouin zone. }
\label{fig1}
\end{figure}
${\bf v}_s^\mu({\bf r})$ satisfy equations similar to (\ref{vs})
with ${\bf R}_i$ replaced by ${\bf R}_i^\mu$ and an overall prefactor $2$. 

It is easy to verify that ${\bf v}_s^\mu$ correspond to zero effective field,
i.e. that $\langle\nabla\times{\bf v}_s^\mu\rangle=0$. 
Evidently, there is no reason to expect LL quantization in 
the system. This property becomes more transparent if we focus on the 
low-energy 
excitations. By linearizing  ${\cal H}_N$ in the vicinity of the four 
nodes as described in Ref.\ \cite{sl1} we obtain ${\cal H}_N\simeq {\cal H}_0+
{\cal H}'$, where
\begin{equation}
{\cal H}_0 =\left( \begin{array}{cc}
v_F\hat p_x & v_\Delta\hat p_y \\
v_\Delta\hat p_y & -v_F\hat p_x
\end{array} \right)
\label{h0}
\end{equation}
is the free Dirac Hamiltonian and 
\begin{equation}
{\cal H}' =m\left( \begin{array}{cc}
v_Fv_{sx}^A & {1\over 2}v_\Delta(v_{sy}^A-v_{sy}^B) \\
{1\over 2}v_\Delta(v_{sy}^A-v_{sy}^B) & v_Fv_{sx}^B
\end{array} \right)
\label{h'}
\end{equation}
is the vector potential term, $v_F$ is the Fermi velocity and 
$v_\Delta=\Delta_0/p_F$ denotes the slope of the gap at the node.  

Our considerations so far have been completely general and apply to arbitrary
distribution of vortices. In the following we illustrate the utility of the 
new Hamiltonian by finding the excitation spectrum in a periodic
square vortex lattice. With minor modifications the same approach can be 
generalized to an arbitrary periodic lattice, such as e.g. triangular. 
We take $A$ and $B$ subsets to coincide with the two sublattices
of the square vortex lattice as illustrated in Figure \ref{fig1}(a). Expanding 
the quasiparticle wavefunction in the plane wave basis 
$\Psi({\bf r})= \sum_{\bf q} \Psi_{\bf q} e^{i{\bf q}\cdot{\bf r}}$ we arrive 
at an equation of the form
\begin{equation}
{\cal H}_0({\bf q})\Psi_{\bf q}+\sum_{\bf K}{\cal H}'({\bf K})\Psi_{{\bf q}+{\bf K}}=E\Psi_{\bf q}.
\label{bloch}
\end{equation}
Because of its periodicity, ${\cal H}'$ only has non-vanishing Fourier 
\begin{figure}[t]
\epsfxsize=8.5cm
\epsffile{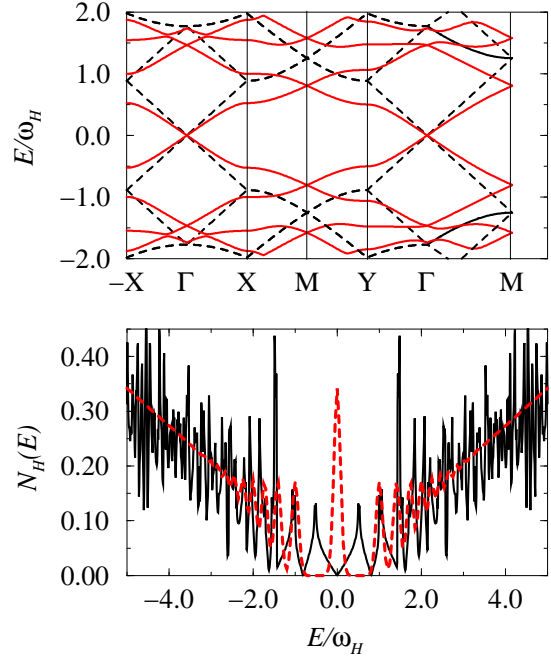}
\caption[]{Upper panel: lowest energy bands near a single node 
for isotropic case $\alpha_D\equiv v_F/v_\Delta=1$.
Dashed lines represent the unperturbed Dirac spectrum of ${\cal H}_0$
while solid lines reflect the effect of the periodic potential ${\cal H}'$.
Lower panel: the corresponding DOS.  Dashed line
represents the thermally broadened DOS resulting from LL 
spectrum of Eq.\ (\ref{gs}).}
\label{fig2}
\end{figure}
components at the reciprocal lattice vectors ${\bf K}={2\pi\over d}(m_x,m_y)$, 
where $(m_x,m_y)$ are integers and 
$d=\sqrt{2\Phi_0/B}$ is the size of the unit cell. 
Aside from the $2\times 2$ matrix structure, Eq.\ (\ref{bloch})
is a standard Bloch equation which we solve by numerical diagonalization
for ${\bf q}$ vectors in the first MBZ, sketched in Figure 
\ref{fig1}(b).

As long as $\lambda\gg d$ the results are independent of $\lambda$ and 
the only free parameter in the low-energy
theory is the Dirac cone anisotropy $\alpha_D=v_F/v_\Delta$. The band 
structure for the isotropic case, $\alpha_D=1$ is presented in Figure
\ref{fig2}. When compared to the unperturbed Dirac bands the periodic 
potential has precisely the expected effect of opening up band gaps at the MBZ
boundaries. The surprising finding is that the magnetic field {\em does not}
destroy the original nodal point, but merely renormalizes the slope
of the dispersion. This finding can be understood as a consequence of the 
exact electron-hole symmetry of the linearized Hamiltonian 
(\ref{h0}-\ref{h'}). The associated 
DOS vanishes at the Fermi level, in contrast
to the peak expected from the LL scenario\cite{gs1,pwa1,janko1} (cf. Figure
\ref{fig2}). The peaks
which appear in DOS are van Hove singularities related to the band
structure and have nothing to do with LL spectrum of Eq.\ (\ref{gs}). 
Observation of these peaks is a challenge to the experimental community. 

\begin{figure}[t]
\epsfxsize=8.5cm
\epsffile{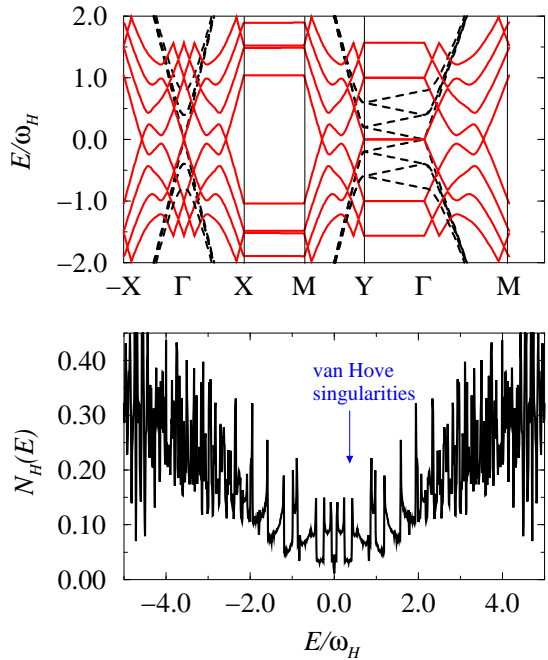}
\caption[]{The same as Figure \ref{fig2} with $\alpha_D=20$. }
\label{fig3}
\end{figure}

In Figure \ref{fig3} we display the band structure for $\alpha_D=20$, 
a value perhaps more relevant for the optimally doped 
YBa$_2$Cu$_3$O$_{7-\delta}$.
 The striking new feature is the formation of additional
{\em lines of nodes} on the Fermi surface [see also Figure \ref{fig1}(b)]
which give rise to a finite DOS at the Fermi surface. This structure can
be understood by considering the $\alpha_D\to\infty$ limit\cite{melnikov1}.
The lines of nodes first appear at $\alpha_D\approx 15$ and with increasing 
nodal anisotropy their number increases. This finding is consistent with 
the prediction of a finite residual DOS based on the semiclassical 
approach\cite{volovik1}, which is expected to be valid when 
$\Delta_0\ll\epsilon_F$, or equivalently $\alpha_D\gg 1$. Qualitatively 
similar results are found for different orientations of the vortex
unit cell with respect to underlying ionic lattice.

In conclusion, we have shown how the collective superfluid response of a 
superconductor
ensures that the effective magnetic field ${\bf B}_{\rm eff}$
seen by a fermionic quasiparticle (and distinct from the physical field 
${\bf B}$) is zero on average, even in the vortex state. 
The physics of a low energy 
quasiparticle in a  superconductor with gap nodes is that of a 
massless Dirac fermion
moving in a vector potential associated with physical supercurrents but zero 
average magnetic field. For a periodic vortex lattice the appropriate 
description is in terms of familiar Bloch waves. A mathematically equivalent
description could be given (in a different gauge) in terms of Landau levels
\cite{gs1,pwa1,janko1} strongly scattered by supercurrents\cite{melnikov1}. 
The fact that 
no trace of LL structure remains in the exact spectrum of excitations 
suggests that the former is a more useful starting point. The LL quantization
remains a domain of relatively high fields\cite{dukan1}. Our conclusions 
are corroborated by the absence of LL spectra in the numerical computations
\cite{wang1} as well as in the experimental tunneling 
data on cuprates\cite{fischer1} and are consistent with scaling arguments  
given previously\cite{volovik1,sl1}.

In the present study we have focused on the leading low-energy, long-wavelength
behavior of the quasiparticles as embodied by the linearized Dirac Hamiltonian
(\ref{h0}-\ref{h'}). In real materials and at higher energies our results
may be modified by the corrections to the linearization, electron-hole
asymmetry, possible internode
scattering, ionic lattice effects and the vortex core physics. These issues,
as well as a more rigorous discussion of the singular gauge transformation, 
are best addressed within the framework of a tight binding calculation to be
reported in a forthcoming publication. 

We emphasize that the central idea of this paper, i.e. that upon proper 
inclusion of the condensate screening the quasiparticles experience effective 
zero average magnetic field, is completely general and robust against any 
effects of short length scale physics. Consequently, our method 
is applicable to any pairing symmetry and arbitrary distribution of
vortices and could be useful 
for understanding the physics of vortex glass and liquid phases, as well as 
the zero-field quantum phase-disordered states such as the nodal 
liquid\cite{balents1}.
We expect that disorder in the vortex positions will smear the
structure apparent in Figures \ref{fig2} and \ref{fig3}, resulting 
in smooth DOS.
Of obvious interest are implications for the 
quasiparticle thermodynamics, transport and localization properties
in statically disordered or fluctuating vortex arrays.

The authors are indebted to A. A. Abrikosov, W. A. Atkinson, A. V. Balatsky,
M. P. A. Fisher, B. Jank\'o, A. H. MacDonald, A. Melikyan,  D. Rainer,
J. A. Sauls, and O. Vafek 
for helpful discussions and to P. W. Anderson, A. S. Melnikov, 
J. R. Schrieffer and G. E. Volovik for correspondence.  This research was 
supported in part by NSF grant DMR-9415549.


\begin{references}
\bibitem{remark4} Here and hereafter we neglect higher derivative terms 
$(\partial_x\partial_y\phi)$ which are unimportant at low energies.
\bibitem{remark2} This is an exact consequence of flux quantization, 
independent on the details of the vortex core physics.
\bibitem{remark1} Note that our result differs from Eq.\ (13a) of 
Ref.\ \cite{pwa1}.
\end{references}
\end{document}